\documentclass[amssymb,aps,epsf,eqsecnum,nofootinbib,showpacs]{revtex4}
\usepackage{graphicx}
\usepackage{pictexwd,dcpic}
\usepackage{wasysym}

\usepackage{makeidx}
\makeindex
\usepackage{hyperref}
\begin{document}

\title{Presymplectic Geometry And The Problem Of Time. Part 2}         
\author{Vasudev~Shyam}
\email{vasudev@cfrce.com}
\author{ B~S~Ramachandra}
\email{bsr@cfrce.com}
\affiliation{Centre for Fundamental Research and Creative Education,\\
Bangalore, India}

\date{\today}          

\begin{abstract}
The Problem of Time in Quantum Gravity is analyzed from a classical presymplectic perspective. In the first part of the paper
the Three Space Approach to General Relativity is introduced via the Barbour-- Foster --\'O Murchadha action and the dynamics
corresponding to a theory of relativistic gravity where 
spacetime is not an a priori requirement. We also look into the nature and physical interpretation of
the constraints in this theory and compare them with those of Standard ADM General Relativity.
We then  study the presymplectic phase space of three space general relativity and discuss briefly the notion of observables and perennials of the system. We then move on to re-deriving the ephemeris lapse identification, and then discuss the notion of re-foliation invariance and its resolution in the conformal theory. Further, we study a new perspective of three space general relativity involving a Hamiltonian reduction of the phase space and subsequently, we discuss the path integral quantization of the same.
\end{abstract}

\maketitle
\section{Introduction}
In a previous paper \cite{vr1}, we developed a framework that could deal with particle models whose dynamics is governed by a completely constrained Hamiltonian. We showed that this need not imply the absence of true physical evolution of the system. In this paper, we apply our framework to Hamiltonian General Relativity wherein the problem of time really originates. In particular, we shall investigate the implications of timelessness in General Relativity for which we shall develop a Hamiltonian framework for the "Relativity Without Relativity" approach pioneered by Barbour et. al. 
\section{The BSW Action and GR Without Spacetime}       
In this section, we take a look into the Barbour--Foster-- \'O Murchadha formalism of geometrodynamics where General Relativity is treated as a theory of dynamical three spaces rather than a theory of three spaces embedded into an a priori assumed Lorentzian four manifold (spacetime).

\subsection{Trimming GR into a Timeless Theory}       
As we have seen in a previous paper\cite{vr1} with a Jacobi type action, we can eliminate time from the action, and replace it with a  reparameterization invariant action that is integrated over an unphysical evolution parameter. We now investigate whether this applies to General Relativity also. We shall begin with the Lagrangian for ADM gravity
\begin{equation} 
S=\int dt d^{3}x N (\sqrt{\gamma}R)-\frac{1}{\sqrt{\gamma}}(K_{ab}K^{ab}-trK^{2})
\end{equation}
We replace $K$ with $k$ where
\begin{equation}
k_{ab}= \dot{\gamma_{ab}}-\mathcal{L}_{\xi ^{a}}\gamma_{ab}
\end{equation}
The $\xi$ is an arbitrary vector field with respect to which the Lie Derivative acting on the metric represents the action of the 3 diffeomorphism group on the configuration space (that turns out to be equal to the shift of ADM gravity) and the over dot denotes differentiation with respect to an unphysical evolution parameter $\lambda $ and so our action now looks like
\begin{equation}
S=\int d\lambda d^{3}x \left[ \frac{1}{4N\sqrt{\gamma}}(k_{ab}k^{ab}-trk^{2})-N\sqrt{\gamma}R \right] \end{equation}
Varying with respect to N, we get
\begin{equation} 
N=\sqrt{\frac{k_{ab}k^{ab}-trk^{2}}{4R}}\end{equation}
 Putting this back into the action we find
 \begin{equation} 
S= \int d\lambda d^{3}x \sqrt{\gamma}\sqrt{R} \sqrt{T} 
\end{equation}
 Here the "Kinetic Energy" term T is
 \begin{equation}
 T= G^{abcd}(\dot{\gamma_{ab}}-\mathcal{L}_{\xi^{a}}\gamma_{ab})(\dot{\gamma_{cb}}-\mathcal{L}_{\xi^{a}}\gamma_{cd}), \end{equation}
and the $G$ is the DeWitt Supermetric. Now the only remaining condition imposed by Barbour et al. to make General Relativity a truly 3 dimensional theory is a consistency condition (constraint propagation) and the relaxation of the embeddability criterion (Dirac Algebra)
\section{The Constraints}
The Hamiltonian of General Relativity, much like those we have considered thus far consists purely of constraints (first class) i.e.
\begin{equation} 
\mathcal{H}=\int d^{3}x (NH + N^{a}H_{a})=C(N)+C(N^{a})
\end{equation}
where
\begin{equation}
 C(N)= \int d^{3}x \left(\frac{1}{\sqrt{\gamma}} G^{abcd}\pi_{ab}\pi_{cd} -\sqrt{\gamma}R \right)
 \end{equation}
and
\begin{equation}
 C(N^{a})=-\int d^{3}x 2N^{a}\nabla_{a}\pi^{ab}
\end{equation}
The Poisson algebra of these constraints is given by the Dirac or the hypersurface deformation algebra
\begin{equation} 
\left\{C(N^{a}),C(N^{b})\right\}=C(\mathcal{L_{N^{a}}}N^ {b})
\end{equation}
\begin{equation}
\left\{C(N^{a}),C(N)\right\}=C(\mathcal{L}_{N^{a}}N) 
\end{equation}
\begin{equation}
\left\{C(N),C(N')\right\}=C(N'\gamma^{ab}\nabla_{b}N-N\gamma^{ab}\nabla_{b}N')
\end{equation}
Here the constraints are to be interpreted as generators of tangential and normal deformations of the three spatial slice embedded into spacetime. That is, the diffeomorphism constraint generates tangential deformations while the Hamiltonian constraint generates the normal deformations. The algebra of the constraints can be seen as the embeddability criterion for the space like slices to be embedded into spacetime. One must note that the Poisson bracket of two Hamiltonian constraints is a diffeomorphism and the algebra does not close to structure constants but instead to phase space dependent functions and so the algebra is not a Lie Algebra. 
\section{The Interpretation of Constraints in the BF\'O Approach} 
In the theory described in the previous section the Hamiltonian constraint of General Relativity arises from a square root identity of the local square root form of the action i.e.
\begin{equation} 
 \frac{1}{\sqrt{\gamma}}(\pi^{ab}\pi_{ab}-\frac{1}{2}tr\pi^{2})-\sqrt{\gamma}R=0 \end{equation}
and the diffeomorphism constraint arises from best matching where
\begin{equation}
\delta_{\xi}S_{BSW}=0 
\end{equation}
that gives us
\begin{equation}
 -2\nabla_{a}\pi^{ab}=0 
\end{equation}
Finally, and most importantly, this approach relaxes the necessity for the Dirac algebra. Instead we have the consistency conditions
\begin{equation}
 \dot{C}\left(\sqrt{\frac{T}{4R}}\right)=0
\end{equation}
and
\begin{equation} 
\dot{C}(\xi^{a})=0 
\end{equation}
This means that the Euler Lagrange equations obtained from varying the BSW action must propagate the constraints. Also, the role of the diffeomorphism constraint remains the same while the Hamiltonian constraint is now a generator of real physical evolution (in accordance with \cite{fos}). With this, the theory is now fully consistent. Our analysis henceforth shall be centered around the canonical phase space of this system.
\section{The Presymplectic Dynamics Of General Relativity}
Now an in depth analysis and characterization of canonical gravity as a three dimensional constrained Hamiltonian theory shall be carried out. We begin by reviewing details about the configuration space and phase space of general relativity.

\subsection{Superspace}
What is the configuration space of General Relativity? Let us answer this question by first looking at what the elements of any configuration space should satisfy. The vacuum Einstein Equations are given by
\begin{equation}
G_{\mu \nu}=0 
\end{equation}
where $G_{\mu \nu}$ is the Einstein Tensor given by
\begin{equation}
 G_{\mu \nu}=R_{\mu \nu}-\frac{1}{2}g_{\mu \nu}R=0 
\end{equation}
Here $ R_{\mu \nu}$ is the Ricci Tensor. Thus the solutions to the Einstein Equations is the set of Lorentzian metrics on spacetime $Met(\mathcal{M})$
The 3+1 formalism makes the identification
\begin{equation} 
\mathcal{M}= R \times \Sigma 
\end{equation}
Consequently, the metrics on $ \Sigma $ too must satisfy the Gauss-Codazzi Equations, which in 3+1 form are written as
\begin{equation}
 -2G_{\mu \nu}n^{\mu}n^{\nu}=^{3}R+tr(K^{2})-(tr(K))^{2}=0 
\end{equation}
\begin{equation}
 -2G_{\mu a}n^{\mu}=^{3}\nabla_{b}K^{a}_{b}-^{3}\nabla_{a}K^{a}_{a}=0
\end{equation}
Here the $K$'s are the extrinsic curvature tensors, and the $n$'s are the normals to $\Sigma$.
(here, $^{3}$ and the Latin indices are written to indicate that these quantities are defined on $\Sigma$). The solution to these equations is given by the three metric $\gamma_{ab}$ and since we are considering the configuration space still, we should note that the velocity analogue for this space is closely related to the extrinsic curvature tensor. It follows directly that the space of solutions should be the space of 3 metrics that obey the Gauss equation and satisfy the Codazzi equation. It is well known that the latter condition implies that they be invariant under 3-Diffeomorphisms. Thus we define Superspace as
\begin{equation} 
\mathcal{S}=\frac{Met(\Sigma)}{Diff(\Sigma)}
\end{equation}
Now care is needed in the definition of the Diffeomorphism group, which we shall restrict to a proper subgroup of the diffeomorphisms that fix a preferred point $\infty \in \Sigma$ such that
\begin{equation} Diff_{F}(\Sigma)=\left\{ \phi \in Diff(\Sigma)| \phi(\infty)=\infty,\phi_{*}(\infty)=Id|_{T_{\infty}\Sigma} \right\}
 \end{equation}
This ensures that the action of this group is free and proper when $\Sigma$ is Closed and Compact and that no metric on $\Sigma$ has a non trivial isometry group, thereby ensuring that
\begin{equation}
\mathcal{S}_{F}=\frac{Met(\Sigma)}{Diff_{F}(\Sigma)}
\end{equation}
is a Manifold. Thus we shift our definition of superspace from $\mathcal{S}$ to $\mathcal{S}_{F}$.
\subsection{Super-Phase Space}
The equations in the previous subsection can be written as
\begin{equation} 
-2G_{\mu \nu}n^{\mu}n^{\nu}=^{3}R+tr(K^{2})-(tr(K))^{2}=H=0 
\end{equation}
\begin{equation}
 -2G_{\mu a}n^{\mu}=^{3}\nabla_{b}K^{a}_{b}-^{3}\nabla_{a}K^{a}_{a}=H_{a}=0
\end{equation}
The data that satisfies this equation is obtained by taking the Legendre transform of the data satisfying the Vacuum Einstein equations on configuration space. So 
\[z^{I}=\left(\begin{array}{cc}\pi^{ab}\\ \gamma_{ab}\end{array}\right)\],
is a point on super phase space and
$$ \pi_{ab}=-\sqrt{\gamma}(K_{ab}-\gamma_{ab}trK),$$
is the momentum conjugate to the metric that satisfies the Hamiltonian constraint. Thus Super Phase space will be the cotangent bundle of Super Space. Now, in the 3 space approach also, we retain this phase space. The constraints are
\[\Phi^{J}[z^{I}]=\left(\begin{array}{cc} \left(\frac{1}{\sqrt{\gamma}} G^{abcd}\pi_{ab}\pi_{cd} -\sqrt{\gamma}R \right) \\ -2\nabla_{a}\pi^{ab} \end{array}\right)=0\].
For our purpose we shall not work with $T^{*}\mathcal{S}_{F}(\Sigma)$ but instead identify $T^{*}Met(\Sigma)$ as the total phase space $\Gamma$. For the explicit mapping between them, see \cite{vm1}.
\section{The Presymplectic Equation}
Let us now look into the Presymplectic dynamics of this system. We begin with out aforementioned phase space $\Gamma$ and on it, we define the Hamiltonian
\begin{equation}
 \mathcal{H}=\int d^{3}x (\frac{N}{\sqrt{\gamma}} G^{abcd}\pi_{ab}\pi_{cd} -N\sqrt{\gamma}R-2\xi^{a}\nabla_{a}\pi^{ab})
\end{equation}
We attain the constraint hypersurface by imposing
\begin{equation}
\frac{\delta \mathcal{H}}{\delta N}|_{\tilde{\Gamma}}=\Phi^{0}[z^{I}]
\end{equation}
\begin{equation}
\frac{\delta \mathcal{H}}{\delta \xi}|_{\tilde{\Gamma}}=\Phi^{1}[z^{I}]
\end{equation}
In totality
$$\Phi^{J}[z^{I}]=0,$$
which gives us the constraint hypersurface. On this hypersurface, we can define the presymplectic form
\begin{equation}\Omega|_{\tilde{\Gamma}}=\int_{\Sigma}d^{3}x \delta\pi_{ab}\land \delta\gamma^{ab} .\end{equation}
Here, $\delta$ is the functional exterior derivative. Henceforth, it shall be used interchangeably with $d_{\delta}$.  We now use the fact that any symplectic vector field on the constraint hypersurface will be locally Hamiltonian for it's flow preserves $\Omega$ i.e. 
$$\mathcal{L}_{X}\Omega|_{\tilde{\Gamma}}=0$$
$$=> (\iota_{X}d_{\delta}\Omega+d_{\delta}\iota_{X}\Omega)|_{\tilde{\Gamma}}=0$$
$$=>(d_{\delta}\iota_{X}\Omega)|_{\tilde{\Gamma}}=0$$
$$=>(\iota_{X}\Omega)|_{\tilde{\Gamma}}=d \mathcal{H}$$
$$i_{*}X=X_{\mathcal{H}}.$$
Here $i$ is the inclusion map from $\tilde{\Gamma}$ to $\Gamma$. From the above calculation we obtain the (locally) Hamiltonian vector field
\begin{equation}X_{\mathcal{H}}=\left[ 2\frac{N}{\sqrt{\gamma}}\left(\pi_{ab}-\frac{1}{2}\gamma_{ab} tr\pi \right)+ \mathcal{L}_{\xi^{a}}\gamma_{ab}\right]\frac{\delta}{\delta \gamma^{ab}}-\end{equation}
\begin{equation}[ N\sqrt{\gamma}\left(R^{ab}-\frac{1}{2}\gamma^{ab}R \right) - \frac{N \gamma^{ab}}{2\sqrt{\gamma}}\left(\pi_{ab}\pi^{ab}-\frac{1}{2}tr\pi^{2}\right)+\frac{2N}{\sqrt{\gamma}}\left(\pi^{ac}\pi^{b}_{c}-\frac{1}{2}\pi^{ab} tr\pi \right)+\end{equation} 
\begin{equation}\sqrt{\gamma}(\nabla^{a}\nabla^{b}N-\gamma^{ab}\nabla^{2}N) + \mathcal{L}_{\xi^{a}}\pi^{ab} ]\frac{\delta}{\delta \pi_{ab}} .\end{equation}
Thus, the presymplectic equation is
\begin{equation}
(X_{\mathcal{H}})^{\flat}|_{\tilde{\Gamma}}=0
\end{equation}
\subsection{Flows And Constraint Propagation}
Since we have determined the dynamical vector field, we can now define the flows of functionals on phase space. We begin with the canonical flow of the phase space variables which is a solution to the Cauchy problem
\begin{equation}
f^{0}_{\mathcal{H}}[z^{I}]=z^{I}
\end{equation}
\begin{equation}
\frac{d}{d \lambda}f^{\lambda}_{\mathcal{H}}[z^{I}]=X_{\mathcal{H}}[z^{I}]
\end{equation}
Now the solution to this is given by
\begin{equation}
 f^{\lambda}_{\mathcal{H}}[z^{I}]=e^{\lambda X_{\mathcal{H}}[z^{I}]}
\end{equation}
\begin{equation}=>\sum_{n=0}^{\infty}\frac{\lambda ^{n}}{n!}X^{n}_{\mathcal{H}}[z^{I}]
\end{equation}
Now, for this formalism to be consistent, we require that the condition
\begin{equation}
f^{\lambda}_{\mathcal{H}}[\Phi^{J}[z^{I}]]=\Phi^{J}[z^{I}]
\end{equation}
should hold, and it does because
\begin{equation}
X_{\mathcal{H}}[H_{a}]=\mathcal{L}_{\vec{\xi}}H_{a}+H\nabla_{a}N \approx 0
\end{equation}
\begin{equation}
X_{\mathcal{H}}[H]=\mathcal{L}_{N}H+N\nabla_{a}H_{a}+2(\nabla^{a}N)H_{a} \approx 0
\end{equation}
($\approx$ denotes weak equivalence, i.e. $f \approx g$ iff $f|_{\tilde{\Gamma}}=g|_{\tilde{\Gamma}}$)\\
and this is because all of the terms in the right hand side are proportional to constraints and their derivatives, all of which vanish weakly (and here they do since we are only considering the flow on the constraint hypersurface). This is nothing but the constraint propagation condition imposed in BF\'O gravity, the difference being that now we require the constraints to be invariant under the flow of the Hamiltonian vector field on the constraint hypersurface rather than being propagated by the Euler Lagrange equations. 
\subsection{Observables and Perennials}
First let us consider a variation on the presymplectic equation previously derived
\begin{equation}
(X_{\mathcal{H}})^{\flat}|_{\tilde{\Gamma}}=0
 \end{equation}
From the presymplectic algorithm we know that $\flat$ is an isomorphism at $\tilde{\Gamma}$, thus the above equation admits the splitting
\begin{equation}
(X_{\mathcal{H}})^{\flat}|_{\tilde{\Gamma}}=(\mathcal{E}_{H(N)})^{\flat}|_{\tilde{\Gamma}}+(\mathcal{G}_{H_{a}(\xi^{a})})^{\flat}|_{\tilde{\Gamma}}=0
\end{equation}
Here, the vector field $\mathcal(E)_{H(N)}$ is the part of the Hamiltonian vector field that generates purely physical evolution
\begin{equation}
\mathcal{E}_{H(N)}|_{\tilde{\Gamma}}= 2\frac{N}{\sqrt{\gamma}}\left(\pi_{ab}-\frac{1}{2}\gamma_{ab} tr\pi \right)\frac{\delta}{\delta \gamma^{ab}}-\end{equation} 
\begin{equation}
[ N\sqrt{\gamma}\left(R^{ab}-\frac{1}{2}\gamma^{ab}R \right) - \frac{N \gamma^{ab}}{2\sqrt{\gamma}}\left(\pi_{ab}\pi^{ab}-\frac{1}{2}tr\pi^{2}\right)+\frac{2N}{\sqrt{\gamma}}\left(\pi^{ac}\pi^{b}_{c}-\frac{1}{2}\pi^{ab} tr\pi \right)+
\end{equation} 
\begin{equation}
\sqrt{\gamma}(\nabla^{a}\nabla^{b}N-\gamma^{ab}\nabla^{2}N)]\frac{\delta}{\delta \pi_{ab}}
\end{equation}
And the generator of Diffeomorphisms or the Gauge Flats
\begin{equation}
\mathcal{G}_{H_{a}(\xi^{a})})|_{\tilde{\Gamma}}=\mathcal{L}_{\xi^{a}}\gamma_{ab}\frac{\delta}{\delta \gamma^{ab}}+\mathcal{L}_{\xi^{a}}\pi^{ab}\frac{\delta}{\delta \pi_{ab}}
\end{equation}
Now we can define the set of Perennials as
\begin{equation}
\mathcal{P}=\mathcal{O}\cap \mathcal{D}=\left\{f[z^{I};\lambda) | X_{\mathcal{H}}[f[z^{I};\lambda)]=0\right\}
\end{equation}
And the set of Observables is defined by the set
\begin{equation}
 \mathcal{O}=\left\{g[z^{I};\lambda)|\mathcal{G}_{H_{a}(\xi^{a})}[g[z^{I};\lambda)]=0\right\}
\end{equation}
and the set $\mathcal{D}$ is defined by
\begin{equation}
\mathcal{D}=\left\{h[z^{I};\lambda)|\mathcal{E}_{H(N)}[h[z^{I};\lambda)]=0\right\}
\end{equation}
For instance the spatial volume of $\Sigma$
$$V=\int\textrm{d}^{3}x \sqrt{\gamma},$$ 
satisfies
$$\mathcal{G}_{H_{a}(\xi^{a})}[V]=\int \textrm{d}^{3}x \mathcal{L}_{\xi^{a}}\gamma^{ab}\frac{\delta \sqrt{\gamma}}{\delta \gamma^{ab}}=(\nabla^{a}\xi_{a}-\nabla^{a}\xi_{a})\sqrt{\gamma}=0.$$
Thus $V \in \mathcal{O}$. Another example of an observable is the ADM mass associated with asymptotically flat spaces.
\subsection{The Ephemeris of BF\'O Gravity }
We can write the evolutionary part of the Hamiltonian vector field as
\begin{equation}
\mathcal{E}_{H(N)}[\cdot]|_{\tilde{\Gamma}}=\left(\frac{\partial}{\partial \lambda}-\mathcal{L_{\vec{\xi}}}\right)[\cdot]
\end{equation}
And now
\begin{equation}
\mathcal{E}_{H(N)}[\gamma_{ab}]|_{\tilde{\Gamma}}=2\frac{N}{\sqrt{\gamma}}(\pi_{ab}-\frac{1}{2}\gamma_{ab}tr\pi )
\end{equation}
We can plug this back into the equation
\begin{equation}
\Phi^{0}[z^{I}]=\left(\frac{1}{\sqrt{\gamma}} G^{abcd}\pi_{ab}\pi_{cd} -\sqrt{\gamma}R \right)=0. \end{equation}
\begin{equation}
=>\left[\frac{1}{4N\sqrt{\gamma}}(G^{abcd}\mathcal{E}_{H(N)}[\gamma_{ab}]\mathcal{E}_{H(N)}[\gamma_{cd}]\right]-N\sqrt{\gamma} R=0
\end{equation}
From this we attain an expression for the Lapse
\begin{equation}N=\sqrt{\frac{G^{abcd}\mathcal{E}_{H(N)}[\gamma_{ab}]\mathcal{E}_{H(N)}[\gamma_{cd}]}{4R}}
\end{equation}
which is exactly the same expression that is derived by Barbour et. al. Now with this Lapse, we can define an ephemeris time by smearing this expression over the evolution parameter
\begin{equation} 
\tau=\mathcal{N}(\lambda,x)=\int \sqrt{\frac{G^{abcd}\mathcal{E}_{H(N)}[\gamma_{ab}]\mathcal{E}_{H(N)}[\gamma_{cd}]}{4R}}d \lambda
\end{equation}
\section{The Persistence of Refoliation Invariance}
We can mark equilocal points on the dynamical histories in phase space due to the temporal metricity of the expression obtained in the previous section. But unlike the particle dynamics theories considered in the previous paper, wherein, even though the actions are dependent on a global parameter, they are locally reparameterization invariant, here, the action remains only globally reparameterization invariant. It is disappointing that we still are not able to rid ourselves of the refoliation invariance in this 3 space approach. To see this let us consider two histories in phase space that are identical upto a point $z_{0}$ and thereafter differ only by a local temporal relabelling, which can occur due to the non global nature of the ephemeris we have obtained. Thus, after $z_{0}$ the phase space curves the three geometries will follow are generated by perhaps $\mathcal{E}_{H(N')}$ and $\mathcal{E}_{H(N'')}$... And so, at a given point $x \in z_{0}$ there will be an equilocal point in the subsequent three geometry $z'$ and $z''$ associated to distinct phase space curves. Thus we are dealing with a theory where the dynamical motions of the system are given by equivalence classes of curves on phase space, which, in configuration space terms would correspond to the sheaves of geodesics that the BSW action generates. In order for this this `hidden' symmetry to be fixed, we shall now look to conformal geometrodynamics wherein we attain a fixed foliation i.e. the CMC(constant mean curvature) foliation. 
In order to find the initial data that satisfy the ADM constraints, York discovered the conformal approach to canonical gravity where he identified conformal three geometries with the true dynamical degrees of freedom of the gravitational field. Thus, the physical gravitational degrees of freedom belong to conformal superspace which is the true configuration space of general relativity. Along similar lines, Barbour et. al. developed a theory of gravitation with the conformal superspace as it's core. In the following subsection we shall briefly discuss their theory.
\subsection{The BSW Action on Conformal Superspace}    
In order to recover GR in the York picture, it is necessary that the metric is not only invariant under conformal transformations, but under those that also preserve the total three dimensional volume of the universe. Thus Barbour et. al. found a BSW action for what they call the CS+V (conformal superspace + volume) theory which is given by
\begin{equation} S= \int \textrm{d}\lambda \textrm{d}^{3}x \hat{\phi}\sqrt{\gamma}\sqrt{R-\frac{8\nabla^{2}\hat{\phi}}{\hat{\phi}}} \sqrt{\hat{T}}, \end{equation}
where $\hat{\phi}$ is the conformal factor, and the metric is `corrected' via $\gamma_{ab}\rightarrow \hat{\phi}^{4}\gamma_{ab}$ in order to account for volume preserving conformal transformations. $\hat{T}$ is given by
$$\hat{T}=\hat{\phi}^{-8}G^{abcd}\frac{d\hat{\phi}^{4}\gamma_{ab}}{d\lambda}\frac{d\hat{\phi}^{4}\gamma_{ab}}{d\lambda}.$$
Then, they find the secondary constraint $$\textrm{tr}p=C,$$ where  $C$ is a spatial constant. This arises due to the free end point variation of the CS+V action. This is nothing but the CMC (constant mean curvature) foliation gauge of GR. On enforcing this constraint, they arrive at the following identity
\begin{equation}\sigma_{ab}\sigma^{ab}-\frac{\textrm{tr}\pi^{2}\hat{\phi}^{12}}{6}-\gamma\hat{\phi}^{8}\left(R-\frac{8\nabla^{2}\hat{\phi}}{\hat{\phi}}\right)=0.\end{equation}
Here $\sigma_{ab}=\pi^{ab}-\frac{1}{3}\gamma^{ab}\textrm{tr}\pi.$ This is the Lichnerowicz York equation. Also, another important result of this theory is the consistency condition which arises on propagating the CMC constraint
\begin{equation}NR-\nabla^{2}N+\frac{N\textrm{tr}\pi^{2}}{4\gamma}=\langle \hat{\phi}^{2}N\left(R-\frac{8\nabla^{2}\hat{\phi}}{\hat{\phi}}\right)+\frac{\hat{\phi}^{6}N\textrm{tr}\pi^{2}}{4\gamma}\rangle.\end{equation}
(for some function $f$ $\langle f \rangle=\frac{\int \textrm{d}^{3}x \hat{\phi}^{6}f}{V}$ where $V$ is the volume of three space)
This is the Lapse Fixing condition.
\subsection{A Unimodular Specialization of the Conformal Factor}
Choosing
$$\hat{\phi}^{4}=\gamma^{-1/3},$$ 
for the conformal factor is more than fruitful. From this choice of the conformal factor, we attain the phase space variables:
$$\bar{\gamma}_{ab}=\gamma^{-1/3}\gamma_{ab},$$
which is unimodular
And
$$\sigma^{ab}=\gamma^{1/3}(\pi^{ab}-\frac{1}{3}\gamma^{ab}\textrm{tr}\pi).$$
Another interesting feature of this system is that it shares the phase space of three space general relativity, $T^{*}Met(\Sigma)$, and the presymplectic potential is given by (See \cite{hc1}, \cite{hco1})
\begin{equation}\int \textrm{d}^{3}x \pi^{ab}\textrm{d}_{\delta}\gamma_{ab}=\int \textrm{d}^{3}x \sigma^{ab}\textrm{d}_{\delta}\bar{\gamma}_{ab}+\textrm{tr}\pi \textrm{d}_{\delta}\textrm{ln}\gamma^{1/3}.\end{equation}
Thus there exists another pair of mutually commuting canonical variables $\textrm{tr}\pi$ and $\textrm{ln}\gamma^{1/3}.$ 
\section{Three Space Gravity on Conformal Superspace}
In this section we are going to describe the dynamics of three space gravity in the variables described in the previous section, and we shall see that, on following the analysis described in the previous sections, we can attain a Hamiltonian reduction of the phase space of three space general relativity and we shall also attempt to quantize the system thus attained. The application of this choice of variables to quantum gravity was first done by Chopin Soo and Hoi Lai Yu in \cite{hc1}.
\subsection{The Constraints}
In the variables described in the previous section, the Hamiltonian becomes
\begin{equation} \mathcal{H}=\int d^{3}x (\frac{N}{\sqrt{\gamma}} \bar{G}_{abcd}\sigma^{ab}\sigma^{cd} -N\sqrt{\gamma}R-N\frac{\textrm{tr}\pi^{2}}{6\sqrt{\gamma}}-2\xi^{a}\nabla_{a}\pi^{ab}). \end{equation}
Here $\bar{G}_{abcd}=\bar{\gamma}_{ac}\bar{\gamma}_{bd}$ is the modified supermetric.
Now, in accordance with our formalism, we attain the constraint surface by imposing
\begin{equation}\frac{\delta \mathcal{H}}{\delta N}|_{\tilde{\Gamma}}=0.\end{equation}
\begin{equation}\frac{\delta \mathcal{H}}{\delta \xi^{a}}|_{\tilde{\Gamma}}=0.\end{equation}
The latter yields the diffeomorphism constraint, but our interest is with the former which gives
\begin{equation}[\bar{G}_{abcd}\sigma^{ab}\sigma^{cd}-\sqrt{\gamma} R-\frac{\textrm{tr}\pi^{2}}{6}]|_{\tilde{\Gamma}}=0.\end{equation}
Therefore, on the constraint submanifold, we find that the Hamiltonian constraint can be reduced into a true Hamiltonian (See \cite{hc1}, \cite{hco1}), i.e.
\begin{equation}-\frac{\textrm{tr}\pi}{\sqrt{6}}=\sqrt{\bar{G}_{abcd}\sigma^{ab}\sigma^{cd}-\gamma R}=\bar{\mathcal{H}}.\end{equation} 
\subsection{The Suspended Hamiltonian Vector Field}
The presymplectic potential is given by
\begin{equation}\Theta|_{\tilde{\Gamma}}=\int \textrm{d}^{3}x [\sigma^{ab}\textrm{d}_{\delta}\bar{\gamma}_{ab}-\bar{\mathcal{H}}\textrm{d}_{\delta}\textrm{ln}\gamma^{1/3}].\end{equation}
It shall suffice to find a `suspended' Hamiltonian vector field which can satisfy the presymplectic equation
\begin{equation}\iota_{\mathcal{X}}\textrm{d}\Theta|_{\tilde{\Gamma}}=0,\end{equation}
or
\begin{equation}(\mathcal{X})^{\flat}|_{\tilde{\Gamma}}=0.\end{equation}
This is given by
\begin{eqnarray}\mathcal{X}=\frac{\delta}{\delta \textrm{ln}\gamma^{1/3}}-[\frac{\sqrt{6}}{2\textrm{tr}\pi}\bar{G}_{abcd}\sigma^{cd}+\mathcal{L}_{\xi^{a}}\bar{\gamma}_{ab}]\frac{\delta}{\delta \bar{\gamma}_{ab}}+[\frac{\sqrt{6}}{2\textrm{tr}\pi}\left(R^{ab}-\frac{1}{3}\gamma^{1/3}\bar{\gamma}^{ab}R \right) \\ +\frac{\sqrt{6}}{2\textrm{tr}\pi}\sigma^{ac}\sigma^{b}_{c}-\mathcal{L}_{\xi^{a}}\sigma^{ab}]\frac{\delta}{\delta \sigma^{ab}}. \end{eqnarray}

\subsection{The Symplectic Realization}
It isn't hard to see that we can write the suspended Hamiltonian vector field as
$$\mathcal{X}=\frac{\delta}{\delta \textrm{ln}\gamma^{1/3}}-X_{\check{\mathcal{H}}}.$$
where $\check{\mathcal{H}}$ is the constrained (but not totally constrained) Hamiltonian
$$\check{\mathcal{H}}=\int \textrm{d}^{3}x -\sqrt{\bar{G}_{abcd}\sigma^{ab}\sigma^{cd}-\gamma R}+\xi^{a}H_{a}$$
Applying this to the Presymplectic equation
$$\iota_{\mathcal{X}}\Omega|_{\tilde{\Gamma}}=0$$
$$=>\int \textrm{d}^{3}x [\iota_{X_{\check{\mathcal{H}}}}\textrm{d}_{\delta}\sigma^{ab}\wedge \textrm{d}_{\delta}\bar{\gamma}_{ab}-\textrm{d}_{\delta}\check{\mathcal{H}}]|_{\tilde{\Gamma}}=0.$$
(Here $\Omega$ is the presymplectic form)
Or
$$(X_{\check{\mathcal{H}}})^{\flat}=\textrm{d}_{\delta}\check{\mathcal{H}},$$
on $\tilde{\Gamma}.$
 The above equation indicates that we can find a symplectic phase space on which $\textrm{d}_{\delta}\sigma^{ab}\wedge \textrm{d}_{\delta}\bar{\gamma}_{ab}$ is the symplectic form. We shall now make this more precise.
One of the key features of a presymplectic manifolds is that the presymplectic form admits a symplectic realization. This implies that the presymplectic form can be expressed as the pullback of a symplectic form. Firstly, there exists a characteristic distribution $\textrm{Ker}\Omega$ (to which $\mathcal{X}$ belongs) on the presymplectic manifold which defines a characteristic foliation on it. This foliation can be described in terms of a fibered manifold $$\Phi:\tilde{\Gamma}\rightarrow \tilde{\Gamma}/\textrm{Ker}\Omega.$$
Now, the symplectic form is given by
$$\omega=\Phi^{*}\Omega.$$
And
$$\omega=\int \textrm{d}^{3}x\textrm{d}_{\delta}\sigma^{ab}\wedge \textrm{d}_{\delta}\bar{\gamma}_{ab}$$
With this, we can summarize the dynamical system we have described in totality via this sequence of maps:
\begin{equation} T^{*}Met{\Sigma}\stackrel{\pi_{\mathcal{H}}^{-1}}\rightarrow \tilde{\Gamma}\stackrel{\Phi}\rightarrow \tilde{\Gamma}/\textrm{Ker}\Omega \end{equation}
Here $\pi_{\mathcal{H}}$ is the sequence of inclusion mappings which define the presymplectic algorithm.
\subsection{The BSW Action}
We shall now derive the BSW action via the canonical Lagrangian approach to this theory. First, the transformations required to go from $\tilde{\Gamma}$ (which we shall write as $\Gamma_{H}$ in this section) to $\Gamma_{E}$ in $TMet(\Sigma)$ :
\[\begindc{0}[70]
\obj(0,1){$T^{*}Met(\Sigma)$}
\obj(2,1){$TMet(\Sigma)$}
\obj(0,0){$\Gamma_{H}$}
\obj(2,0){$\Gamma_{E}$}
\mor{$TMet(\Sigma)$}{$T^{*}Met(\Sigma)$}{$FL$}[-1,0]
\mor{$\Gamma_{H}$}{$T^{*}Met(\Sigma)$}{$\pi_{\mathcal{H}}$}
\mor{$\Gamma_{E}$}{$TMet(\Sigma)$}{$\pi_{E}$}[-1,0]
\mor{$\Gamma_{H}$}{$\Gamma_{E}$}{$\pi_{E}^{-1}\circ (FL)^{-1}\circ \pi_{\mathcal{H}}$}[-1,0]
\enddc\]
Now, 
\[\begindc{0}[70]
\obj(0,0){$\Gamma_{H}$}
\obj(2,0){$\Gamma_{E}$}
\obj(0,-1){$\Gamma_{H}/Ker\Omega$}
\mor{$\Gamma_{H}$}{$\Gamma_{E}$}{$\pi_{E}^{-1}\circ (FL)^{-1}\circ \pi_{\mathcal{H}}$}[1,0]
\mor{$\Gamma_{H}$}{$\Gamma_{H}/Ker\Omega$}{$\Phi$}[-1,0]
\mor{$\Gamma_{H}/Ker\Omega$}{$\Gamma_{E}$}{$\Phi^{-1}\circ(\pi_{E}^{-1}\circ (FL)^{-1}\circ \pi_{\mathcal{H}})$}[-1,0]
\enddc\]
We shall refer to the map $\Phi^{-1}\circ(\pi_{E}^{-1}\circ (FL)^{-1}\circ \pi_{\mathcal{H}})$ as $\mathfrak{J}$ for brevity. 
With this, we shall move on to the derivation of the BSW action.
The Hamiltonian vector field on $\Gamma_{H}/\textrm{Ker}\Omega$ must satisfy
$$\mathcal{L}_{X_{\bar{\mathcal{H}}}}\omega=0,$$
$$=>\mathcal{L}_{X_{\bar{\mathcal{H}}}}\textrm{d}_{\delta}\theta=\textrm{d}_{\delta}\mathcal{L}_{X_{\bar{\mathcal{H}}}}\theta=0.$$
By Poincare Lemma
$$\mathcal{L}_{X_{\bar{\mathcal{H}}}}\theta=\textrm{d}_{\delta}\mathcal{A}$$
$$=>\mathcal{L}_{X_{\bar{\mathcal{H}}}}\Phi^{*}\Theta=\Phi^{*}\mathcal{L}_{X_{\bar{\mathcal{H}}}}\Theta=\textrm{d}_{\delta}\mathcal{A}$$
$$=>\Phi^{*}(\iota_{X_{\bar{\mathcal{H}}}}\textrm{d}_{\delta}\Theta+\textrm{d}_{\delta}\iota_{X_{\bar{\mathcal{H}}}}\Theta)=\textrm{d}_{\delta}\mathcal{A}.$$
Because $\Theta$ is presymplectic,
$$\Phi^{*}\textrm{d}_{\delta}\iota_{X_{\bar{\mathcal{H}}}}\Theta=\textrm{d}_{\delta}\mathcal{A}$$
$$=>\Phi^{*}\iota_{X_{\bar{\mathcal{H}}}}\Theta=\mathcal{A}.$$
Using the map previously derived and the form of the evolutionary vector field in equation (21) we find that the BSW action is given by
\begin{equation}\mathfrak{J}\mathcal{A}=\int \textrm{d}^{3}x \textrm{d}\lambda \sqrt{\gamma R}\sqrt{(\textrm{ln}\dot{\gamma}^{1/3}-\mathcal{L}_{\xi^{a}}\textrm{ln}\gamma^{1/3})^{2}-\bar{G}^{abcd}(\dot{\bar{\gamma}}_{ab}-\mathcal{L}_{\xi^{a}}\bar{\gamma}_{ab})(\dot{\bar{\gamma}}_{cd}-\mathcal{L}_{\xi^{a}}\bar{\gamma}_{cd})}.\end{equation}
The overdot refers to differentiation w.r.t the evolution parameter.  The generalized version of this action was derived in \cite{hc1}. \\

On applying the best matching procedure, this action will yield the diffeomorphism constraint, but more importantly, the geodesics defined by the Euler Lagrange equations for this action shall completely determine the dynamics of general relativity on conformal superspace.
\section{Path Integral Quantization}
In this section, we shall attempt path integral quantization of this theory. In order to do so we shall treat the reparameterization invariance of $\lambda$ as a gauge invariance and gauge fix it. The quantum `time' shall then be $\textrm{ln}\gamma^{1/3}.$ Also, the 3-diffeomorphism invariance has to be gauge fixed.

\subsection{Gauge Fixing}
At the very beginning of this series of papers, we attempted to find a mathematical framework which allows us to deal with the reparameterization invariance of the evolution parameter. This invariance is at the very heart of the notion of timelessness. Now we shall treat it as a gauge invariance and gauge fix it in order to quantize this theory. We know that
\begin{equation}\frac{\textrm{tr}\pi}{\sqrt{6}}=-\sqrt{\bar{G}_{abcd}\sigma^{ab}\sigma^{cd}-\gamma R}=\bar{\mathcal{H}}.\end{equation}
Now,
$$[\frac{\textrm{tr}\pi}{\sqrt{6}}+\sqrt{\bar{G}_{abcd}\sigma^{ab}\sigma^{cd}-\gamma R}]=\bar{\mathcal{H}}',$$
 is the constraint which generates reparmeterizations.
The gauge fixing here is the condition 
$$[\tau-\lambda]=0.$$
Here $\tau=\textrm{ln}\gamma^{1/3}.$ This must intersect every gauge orbit in exactly one point.
The mathematical implication of this condition lies in the Fadeev Popov determinant
$$\textrm{det}|X_{\bar{\mathcal{H}}'}[\tau-\lambda]|\neq 0$$
In order to see if it holds, first we calculate
$$X_{\bar{\mathcal{H}}'}[\tau-\lambda]=\frac{1}{3\sqrt{\gamma}}\textrm{tr}\pi,$$
therefore
$$\textrm{det}|X_{\bar{\mathcal{H}}'}[\tau-\lambda]|\neq 0$$
Thus, $\textrm{ln}\gamma^{1/3}$ can be used as the time function.
In the quantum theory, we can impose the diffeomorphism constraint via
$$\nabla_{a}\left\{\frac{\delta \Psi}{\delta \bar{\gamma}_{ab}}\right\}=0.$$
Or, in the path integral we can impose a gauge fixing condition
$$\dot{\xi}^{a}=\chi^{a}.$$
And, we have the Faddeev Popov determinant
$$\textrm{det}|\mathcal{G}_{\xi^{a}}[\dot{\xi}^{a}-\chi^{a}]|\neq 0$$
\subsection{The Path Integral}
The gauge fixed path integral is given by
$$Z=\int \mathcal{D}\bar{\gamma}_{ab}\mathcal{D}\sigma^{ab}\mathcal{D}\tau \mathcal{D}\chi^{a} \mathcal{D}\frac{\textrm{tr}\pi}{\sqrt{6}}\mathcal{D}\bar{\mathcal{H}}'\mathcal{D}H_{a}\delta(\tau-\lambda)\textrm{det}|X_{\bar{\mathcal{H}}'}[\tau-\lambda]|$$ $$\delta(\dot{\xi}^{a}-\chi^{a})\textrm{det}|\mathcal{G}_{\xi^{a}}[\dot{\xi}^{a}-\chi^{a}|e^{\left\{i\int \textrm{d}\lambda\int\textrm{d}^{3}x (\dot{\bar{\gamma}}_{ab}\sigma^{ab}+\dot{\tau}\frac{\textrm{tr}\pi}{\sqrt{6}}-\dot{\xi}^{a}H_{a})\right\}}.$$

On integrating over all the gauge variables and the gauge fixings we get the phase space path integral given by
$$Z[\bar{\gamma}_{ab},\sigma^{ab}]=\int \mathcal{D}\bar{\gamma}_{ab} \mathcal{D}\sigma^{ab} e^{i\left\{ \int \textrm{d}\tau\textrm{d}^{3}x(\sigma^{ab} \frac{\delta \bar{\gamma}_{ab}}{\delta \tau}-\bar{\mathcal{H}})-\dot{\xi}^{a}H_{a}\right\}}$$
This path integral satisfies the Schroedinger equation
\begin{equation}i\frac{\delta \Psi}{\delta \tau}=\hat{\bar{\mathcal{H}}}\Psi=\frac{\textrm{tr}\pi}{\sqrt{6}}\Psi \end{equation}
Where $\Psi$ is the quantum wave function of the universe. In the metric representation, we also have the equation for the momentum conjugate
$$\hat{\sigma}^{ab}\Psi=-i\bar{\delta}^{cd}_{ab}\frac{\delta}{\delta \bar{\gamma}_{ab}}\Psi.$$
(Here $\bar{\delta}^{ab}_{cd}=\frac{1}{2}\delta^{(a}_{c}\delta^{b)}_{d}-\frac{1}{3}\bar{\gamma}^{ab}\bar{\gamma}_{cd}$).
Along with this, we have the canonical commutation relation 
$$[\hat{\bar{\gamma}}_{ab}(x),\hat{\sigma}^{cd}(y)]=i\bar{\delta}^{cd}_{ab}\delta(x-y).$$
 \\
We shall now obtain the semi classical limit of this theory by substituting $\Psi=e^{i\mathcal{S}}$ into the Schroedinger equation, we get
$$\frac{\delta \mathcal{S}}{\delta \tau}-\bar{\mathcal{H}}\left(\bar{\gamma}_{ab},\bar{\delta}^{ab}_{cd}\frac{\delta \mathcal{S}}{\delta \bar{\gamma}_{cd}}\right)=0.$$
Applying the separation ansatz
$$\mathcal{S}=\mathcal{W}+\frac{\textrm{tr}\pi}{\sqrt{6}}\tau,$$
we arrive at the equation
$$\sqrt{\bar{G}^{abcd}\frac{\delta \mathcal{W}}{
\delta\bar{\gamma}_{ab}}\frac{\delta \mathcal{W}}{\delta \bar{\gamma}_{cd}}-\gamma R}=-\frac{\textrm{tr}\pi}{\sqrt{6}}.$$
In phase space co ordinates, this is just
$$\bar{\mathcal{H}}=-\frac{\textrm{tr}\pi}{\sqrt{6}}.$$
Thus we have the quantum theory.
\section{Concluding Remarks}
Thus we have found a description of three space general relativity where the dynamics is governed by an effective constrained, but not totally constrained Hamiltonian. Unlike in conventional canonical gravity, the reduction of the extended phase space does not trivialize the dynamics of the system. This is thanks to the intrinsic time which assures the deparameterization of the system. Although we have shown a quantization of this system, it is purely at the formal level, a more precise approach via geometric quantization shall be the subject of future papers.
\begin{acknowledgments}
We would like to thank Julian Barbour for valuable suggestions to a related part of this work recorded in another paper. This work was carried out at the \textit{Centre for Fundamental Research and Creative Education}, Bangalore, India. We would like to acknowledge the Director Ms. Pratiti B R for facilitating an atmosphere of free scientific inquiry so conducive to creativity. We would also like to thank our fellow researchers Magnona H Shastry, Madhavan Venkatesh, Karthik T Vasu and Arvind Dudi.

\end{acknowledgments}
 
\end{document}